\documentclass[twocolumn,letter]{jpsj2} 
\title{Anomalous Pressure Dependence of the Superconducting Transition Temperature in 
FeSe$_{\rm 1-x}$ Studied by DC Magnetic Measurements}

\author{Kiyotaka \textsc{Miyoshi}, Yuta \textsc{Takaichi}, Eriko \textsc{Mutou}, Kenji \textsc{Fujiwara}, and Jun \textsc{Takeuchi}} 

\inst{%
Department of Material Science, Shimane University, Matsue 690-8504
}

\abst{The pressure dependence of superconducting transition temperature $T_{\rm c}$ 
has been investigated through the DC magnetic measurements for 
FeSe$_{0.8}$ and FeSe$_{1.0}$. For both samples, 
with increasing pressure $P$, the $T_{\rm c}$$-$$P$ curve exhibits 
a two-step increase, showing a local maximum of $\sim$11 K at $P$$\sim$1.0 GPa 
and a rapid increase with an extremely large pressure coefficient 
for $P$$>$1.5 GPa. $T_{\rm c}$ saturates at $\sim$25 K (21 K) 
in FeSe$_{1.0}$ (FeSe$_{0.8}$) for $P$$>$3 GPa. 
A rapid decrease in superconducting volume fraction is observed 
with an increase in $T_{\rm c}$ above 1.5 GPa, 
suggesting the presence of electronic inhomogeneity. 
}

\kword{FeSe, high pressure, superconductivity, DC magnetization}

\begin{document}
\maketitle

Since the discovery of superconductivity in LaFeAsO$_{\rm 1-x}$F$_{\rm x}$ ($T_{\rm c}$=26 K)\cite{kamihara}, 
a great deal of effort has been devoted to explore new superconductors in the related compounds, 
leading to a rich variety of Fe-based superconductors, such as Ba$_{\rm 1-x}$K$_{\rm x}$Fe$_2$As$_2$\cite{rotter,sasmal}, 
Li$_{\rm 1-x}$FeAs\cite{tapp,wang} and FeSe\cite{hsu}, and an increase in $T_{\rm c}$ up to 55 K, 
which is demonstrated by replacing La with Sm in LaFeAsO$_{\rm 1-x}$F$_{\rm x}$\cite{chen}. 
Indeed, the appearance of superconductivity in Fe-based compounds is intriguing because 
these compounds are mostly magnetic. The superconducting phase is therefore 
in the vicinity of the magnetic phase suggesting the possible pairing mechanism of the magnetic origin. 

The application of pressure also increases $T_{\rm c}$ in LaFeAsO$_{\rm 1-x}$F$_{\rm x}$. 
The measurements of electrical resistivity ($\rho$) under pressure have been performed by 
Takahashi $et$ $al$., suggesting that $T_{\rm c}$ increases with increasing pressure $P$ and 
shows a maximum of 43 K at $P$$\sim$4 GPa\cite{takahashi,okada}. 
In contrast, $T_{\rm c}$ for NdFeAsO$_{\rm 1-\delta}$ has 
been found to decrease monotonically from 54 to 16 K 
by the application of pressure up to 18 GPa\cite{takeshita}.  
The rapid increase in $T_{\rm c}$ under pressure in LaFeAsO$_{\rm 1-x}$F$_{\rm x}$ is expected, considering 
that SmFeAsO$_{\rm 1-x}$F$_{\rm x}$ with a higher $T_{\rm c}$ of 55 K has smaller lattice constants. 
It is somewhat surprising that 
the maximum of $T_{\rm c}$ for LaFeAsO$_{\rm 1-x}$F$_{\rm x}$ is fairly lower than 55 K. 
In SrFe$_2$As$_2$, pressure-induced superconductivity has been found through the measurements 
of $\rho$($T$)\cite{kotegawa,igawa}. 
The application of pressure affects $T_{\rm c}$ strongly because it
gives rise to structural modulations in the compound. 
It is important to elucidate the relationship between $T_{\rm c}$ and the local lattice deformation under pressure 
to gain more insight into the mechanism of superconductivity. 

FeSe has a PbO-type crystal structure composed of the stacking layers of Fe$_2$Se$_2$, which is 
analogous to the Fe$_2$As$_2$ layers commonly contained 
in iron-arsenide-based superconductors. The occurrence of superconductivity below $T_{\rm c}$$\sim$8 K has been 
reported first by Hsu $et$ $al$. for FeSe$_{0.88}$\cite{hsu}. 
Fang $et$ $al$. have studied the superconductivity of Fe(Se$_{\rm 1-x}$Te$_{\rm x}$)$_{0.82}$ and 
found that Fe(Se$_{0.4}$Te$_{0.6}$)$_{0.82}$ has a $T_{\rm c}$ of $\sim$14 K\cite{fang}. 
Moreover, Mizuguchi $et$ $al$. have found that the application of pressure increases the $T_{\rm c}$ of FeSe in the 
$\rho$($T$) measurements\cite{mizutani}. They suggest that $T_{\rm c}$ is 27 K at $P$=1.48 GPa, regarding the 
onset temperature of resistivity drop as $T_{\rm c}$. Recently, Medvedev $et$ $al$. have performed 
the $\rho$($T$) measurements under high pressure, suggesting that $T_{\rm c}$ shows a broad 
maximum of $\sim$37 K at $P$$\sim$9 GPa\cite{medvedev}. They also adopt the onset of resistivity drop as $T_{\rm c}$. 
There are some other groups who have reported the $T_{\rm c}$$-$$P$ relation 
determined from the onset of resistivity drop\cite{margadonna,garbarino}. 
In contrast, Masaki $et$ $al$. have defined $T_{\rm c}$ by zero resistance temperature 
and suggested that the pressure dependence of $T_{\rm c}$ 
shows a plateau between $P$=0.5$-$1.5 GPa and a broad maximum of 21 K at $P$$\sim$3.5 GPa in FeSe$_{0.92}$\cite{masaki}. 
It is a difficult problem to determine $T_{\rm c}$ from the resistivity data 
when the resistivity drop occurs over a wide temperature range under pressure. 
The pressure dependence of $T_{\rm c}$ in FeSe is currently a subject of controversy. 

In the present work, we have performed DC magnetization measurements for FeSe$_{\rm 1-x}$ ($x$=0 and 0.2) 
under pressure using a diamond anvil cell (DAC) to determine the $T_{\rm c}$$-$$P$ relation precisely.  
Our DC magnetic measurement using DAC is a powerful technique to establish the $T_{\rm c}$$-$$P$ relation, 
since $T_{\rm c}$ is uniquely determined from the diamagnetic onset temperature. 
For AOs$_2$O$_6$ (A=K, Rb, Cs) pyrochlore superconductors, 
the characteristic $T_{\rm c}$$-$$P$ relations have been successfully revealed by the measurements\cite{miyoshi}. 
In this Letter, it has been found that the $T_{\rm c}$$-$$P$ curve is nonmonotonic and 
shows a maximum at $P$$\sim$1.0 GPa and an abrupt increase for $P$$>$1.5 GPa 
followed by the saturation at $T_{\rm c}$$\sim$25 K (21 K) for $P$$>$3 GPa in FeSe$_{1.0}$ (FeSe$_{0.8}$). 
The $T_{\rm c}$$-$$P$ curves are qualitatively similar to that determined by Masaki $et$ $al$\cite{masaki}.  

The polycrystalline samples with nominal compositions FeSe$_{\rm 1-x}$ ($x$=0 and 0.2) 
used in this study were synthesized by a solid-state reaction technique 
similar to that described in the literature.\cite{hsu,fang,mizutani} 
$T_{\rm c}$ at ambient pressure determined from the diamagnetic onset was $\sim$7 K for both samples. 
For the magnetic measurements under high pressure, a miniature DAC with an outer 
diameter of 8 mm was used to generate high pressure and 
combined with a sample rod of a commercial SQUID magnetometer. 
The details of the DAC are given elsewhere.\cite{mito}
The FeSe$_{\rm 1-x}$ sample was loaded into the gasket hole together 
with a small piece of high-purity lead (Pb) to realize the $in$ $situ$ observation of pressure
by determining the pressure from the $T_{\rm c}$ shift of Pb. 
Magnetization data for the small amounts of FeSe$_{\rm 1-x}$ and Pb were obtained by subtracting the 
magnetic contribution of DAC measured in an empty run from the total magnetization data. 
Most of the measurements have been done by using Daphne oil 7373 as a pressure transmitting medium. 
Daphne oil 7474 that solidifies at $P_{\rm c}$$\sim$3.7 GPa at room temperature, which is higher than $P_{\rm c}$ 
for Daphne 7373 ($\sim$2.2 GPa), was also used. 
\begin{figure}[t]
\begin{center}
\includegraphics[width=8cm]{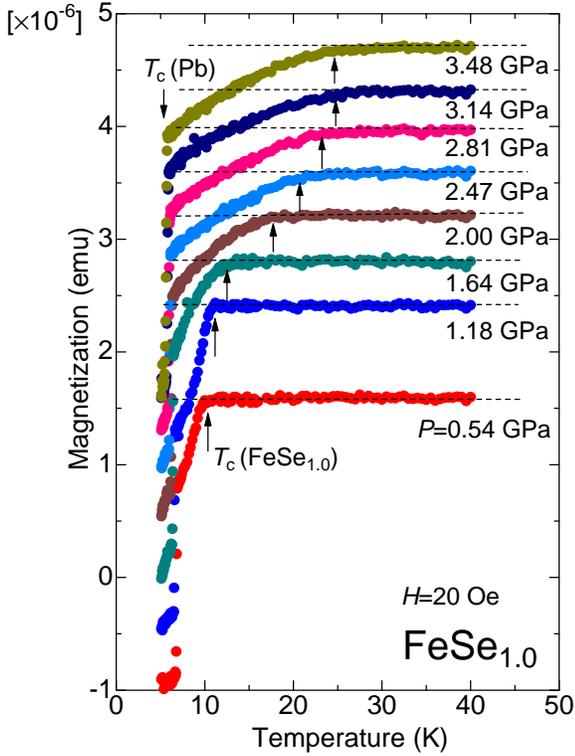}
\end{center}
\caption{(Color on line) Temperature dependence of zero-field-cooled DC magnetization 
for FeSe$_{1.0}$ under various pressures. 
The data are intentionally shifted along the vertical axis for clarity. 
}
\label{f1}
\end{figure}

We show the temperature variations of zero-field-cooled DC magnetization $M$ 
measured with a magnetic field of $H$=20 Oe for FeSe$_{1.0}$ under various pressures in Fig. 1.  
In the figure, the $M$$-$$T$ curve at $P$=0.54 GPa exhibits a sudden decrease at $\sim$10 K, 
indicating the superconducting transition of FeSe$_{1.0}$. 
In addition, a sharp decrease at $\sim$7 K that corresponds to the diamagnetic onset of Pb 
for pressure calibration is observed. A sharp diamagnetic response of FeSe$_{1.0}$ is also observed at $P$=1.18 GPa, 
indicating $T_{\rm c}$$\sim$11 K, slightly higher than $T_{\rm c}$ at $P$=0.54 GPa. At $P$=1.64 GPa, 
$T_{\rm c}$ is further increased to 12.5 K, but the superconducting transition appears to lose its sharpness. 
On further application of pressure up to $P$=3.14 GPa, $T_{\rm c}$ increases very rapidly and reaches 25 K, 
although the transition becomes even broader. The $M$$-$$T$ curve at $P$=3.48 GPa also 
indicates $T_{\rm c}$$\sim$25 K, suggesting that $T_{\rm c}$ is independent of pressure above $P$$\sim$3 GPa. 

To clarify the pressure variation in $T_{\rm c}$ in FeSe$_{1.0}$ in detail, 
the $M$$-$$T$ data under pressure 
were collected by changing pressure in small steps. The results for $P$$<$2 GPa are shown in Fig. 2. 
In the figure, it is found that $T_{\rm c}$ at $P$=0.82 GPa is $\sim$11 K, which is 
higher than $T_{\rm c}$ at $P$=0.42 GPa. 
However, $T_{\rm c}$ gradually decreases to $\sim$10.5 K as pressure increases from 0.82 to 1.42 GPa, 
suggesting that $T_{\rm c}$ exhibits a local maximum near $P$$\sim$0.82 GPa. 
Above $P$=1.42 GPa, a rapid increase in $T_{\rm c}$ is observed. 
In the inset of Fig. 2, the $M$$-$$T$ curves obtained at higher pressures up to $P$=4.78 GPa are shown. 
As seen in the inset, a rapid increase in $T_{\rm c}$ is also seen above $P$=1.90 GPa, 
and the saturation of $T_{\rm c}$ is marked for $P$$\geq$2.67 GPa. 

We have also performed $M$$-$$T$ measurements at various pressures for FeSe$_{0.8}$ 
in order to investigate the sample dependence of the $T_{\rm c}$$-$$P$ relation. 
Figure 3 shows the results of the measurements. 
In Fig. 3, it is found that the $M$$-$$T$ curve at $P$=0.96 GPa indicates $T_{\rm c}$$\sim$11.5 K, 
which is the highest $T_{\rm c}$ for $P$$\leq$1.61 GPa. 
This suggests a local maximum of $T_{\rm c}$ at $P$=0.8$-$1 GPa, similar to the case of FeSe$_{1.0}$. 
Furthermore, a rapid increase in $T_{\rm c}$ is demonstrated in the $M$$-$$T$ curves above $P$=1.61 GPa. 
The $M$$-$$T$ curves above $P$=2.35 GPa suggest a saturation tendency 
of $T_{\rm c}$. The $M$$-$$T$ curves are also shown in the inset of Fig. 3, 
where $T_{\rm c}$ is found to take a constant value of $\sim$21 K for $P$$\geq$3.6 GPa. 
We should note that the slope of the $M$$-$$T$ curve below $T_{\rm c}$ and the amplitude of 
magnetization at 5 K decrease significantly with increasing pressure up to 3.54 GPa, 
indicating a decrease in superconducting volume fraction under high pressure. 
The behavior appears above $P$=1.61 GPa, similar to that observed in FeSe$_{1.0}$.  
In Fig. 3, the volume fraction $p$($\frac12$$T_{\rm c}$) at $P$=3.54 GPa is estimated to be 
$\sim$1$/$4 of that at $P$=0.62 GPa by comparing the amplitude of magnetization at $T$=$\frac12$$T_{\rm c}$. 
For $P$$>$2.5 GPa, the volume fraction appears to be pressure independent. 
\begin{figure}[t]
\begin{center}
\includegraphics[width=8cm]{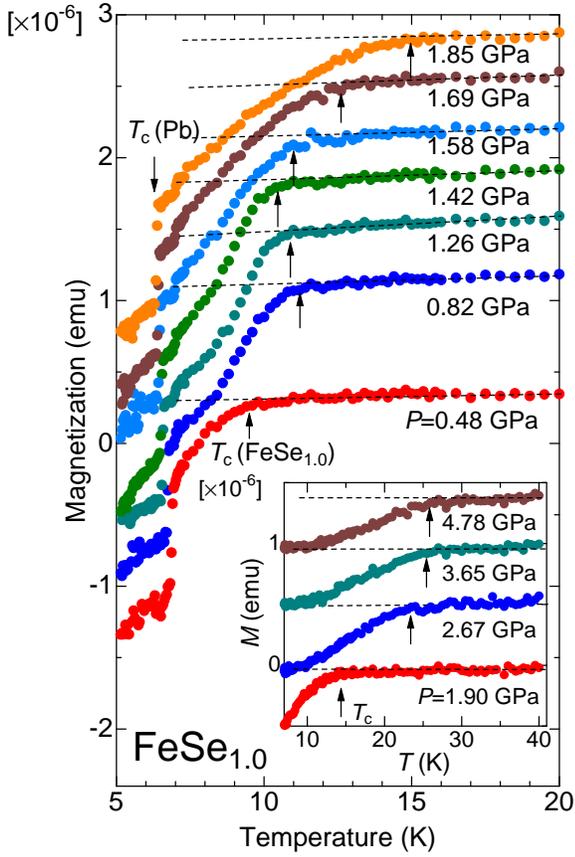}
\end{center}
\caption{(Color on line) Zero-field-cooled DC magnetization versus temperature curves 
for FeSe$_{1.0}$ at pressures up to 2 GPa. The inset shows those obtained at higher pressure. 
The data are intentionally shifted along the vertical axis for clarity. 
}
\label{f1}
\end{figure}
\begin{figure}[t]
\begin{center}
\includegraphics[width=8cm]{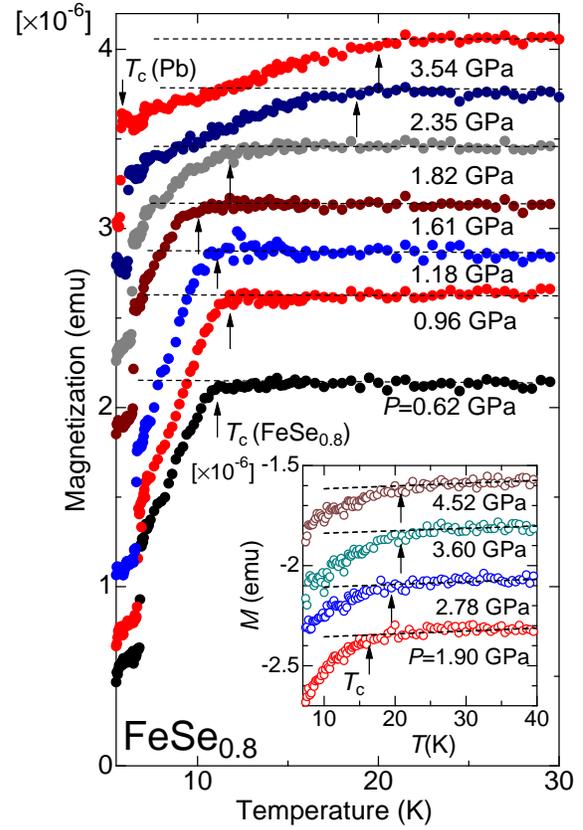}
\end{center}
\caption{(Color on line) Zero-field-cooled DC magnetization versus temperature curves 
for FeSe$_{0.8}$ at various pressures. The inset shows those obtained in the second run. 
The data are intentionally shifted along the vertical axis for clarity. 
}
\label{f1}
\end{figure}

$T_{\rm c}$ for FeSe$_{1.0}$ and FeSe$_{0.8}$ 
is plotted as a function of pressure in Figs. 4(a) and 4(b), respectively. 
For both samples, it is found that the $T_{\rm c}$$-$$P$ curve exhibits a maximum of $\sim$11 K 
at $P$$\sim$1.0 GPa. By the further application of pressure, 
$T_{\rm c}$ is found to increase rapidly with an extremely large pressure 
coefficient of d$T_{\rm c}$$/$d$P$$\sim$10 K/GPa after the sharp upturn at $P$$\sim$1.5 GPa. 
Above $P$$\sim$3 GPa, $T_{\rm c}$ for FeSe$_{1.0}$ saturates and takes a constant value of 
$\sim$25 K. In FeSe$_{0.8}$, $T_{\rm c}$ is also pressure independent above $P$$\sim$3 GPa, reaching 21 K. 
The $T_{\rm c}$$-$$P$ curve for FeSe$_{0.8}$ is almost identical to that determined from 
the zero resistive temperature for FeSe$_{0.92}$ by Masaki $et$ $al$.\cite{masaki}, 
but these curves are different from those determined 
from the onset of resistivity drop\cite{medvedev,margadonna,garbarino}. 
Thus, the diamagnetic onset agrees with the zero resistive temperature 
in the FeSe superconductor for a wide pressure range, and they are considered as reliable markers of $T_{\rm c}$. 
It should be noted that $T_{\rm cm}$, the maximum of $T_{\rm c}$ for $P$$>$3.0 GPa, is $\sim$25 K for FeSe$_{1.0}$ 
but it is $\sim$21 K for FeSe$_{0.8}$ and FeSe$_{0.92}$\cite{masaki}. 
$T_{\rm cm}$ appears to depend on the presence of Se deficit. 
However, the composition $x$ of our specimens is nominal and the exact value is unknown. 
Recently, McQueen $et$ $al$. have prepared the specimens from iron pieces and selenium shot so as to prevent the formation of 
spurious oxides and defects, and showed that Fe$_{1.01}$Se (FeSe$_{0.99}$) and Fe$_{1.02}$Se (FeSe$_{0.98}$) are 
superconducting but Fe$_{1.03}$Se (FeSe$_{0.97}$) is not, suggesting that 
the superconductivity of FeSe is destroyed by very small changes in stoichiometry\cite{mcqeen}. 
They have also revealed that a sample prepared from Fe powder starting with a significant iron excess 
contains secondary phases of Fe and Fe$_{3}$O$_{4}$ and the main phase of stoichiometric FeSe. 
Thus, the FeSe$_{0.8}$ samples used in this study are considered 
to contain an almost stoichiometric FeSe phase as well as FeSe$_{1.0}$. 
Also, FeSe$_{0.92}$ used in the $\rho$($T$) measurements would 
contain an almost stoichiometric FeSe phase. 
The origin of the difference in $T_{\rm cm}$ is unclear, but 
we could not exclude the possibility that other specimens of slightly different composition 
exhibit a $T_{\rm cm}$ more than 25 K. 
\begin{figure}[t]
\begin{center}
\includegraphics[width=8.5cm]{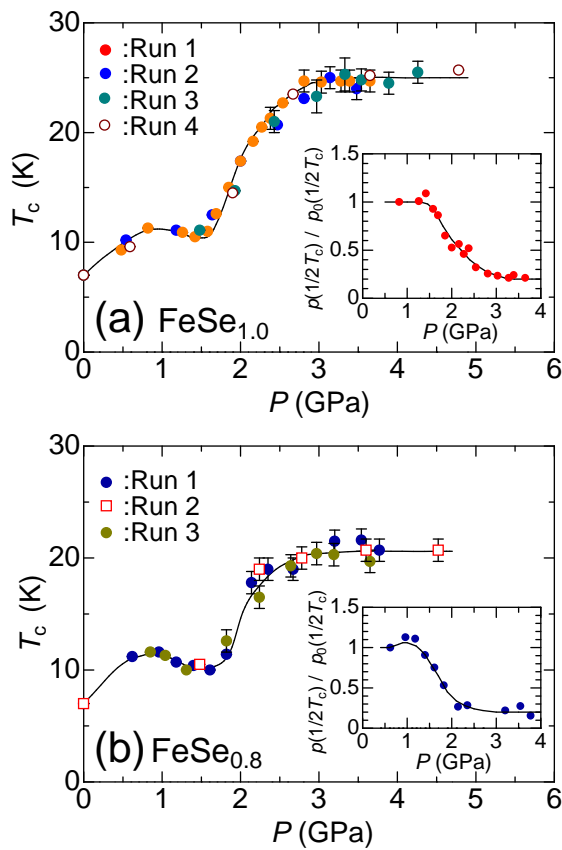}
\end{center}
\caption{(Color on line) Pressure ($P$) dependence of superconducting transition temperature $T_{\rm c}$ for 
FeSe$_{1.0}$ (a) and FeSe$_{0.8}$ (b). The inset shows plots of 
superconducting volume fraction ($p$) at $T$=$\frac12$$T_{\rm c}$ normalized to that at low pressure ($p_0$) 
versus pressure for FeSe$_{1.0}$ (a) and FeSe$_{0.8}$ (b). The solid lines are guides for the eyes. 
}
\label{f1}
\end{figure}

In the insets of Figs. 4(a) and 4(b), the superconducting volume fraction $p$ at $T$=$\frac12$$T_{\rm c}$ 
normalized to that at low pressure $p_0$($\frac12$$T_{\rm c}$) is plotted as a function of pressure for FeSe$_{1.0}$ and FeSe$_{0.8}$, respectively. 
For both samples, $p$($\frac12$$T_{\rm c}$)$/$$p_0$($\frac12$$T_{\rm c}$) is found to decrease rapidly 
above $\sim$1.5 GPa and shows a constant value of $\sim$0.2 above 2.5$-$3.0 GPa. 
This indicates that the volume fraction of superconductivity at $\frac12$$T_{\rm c}$ is 
reduced to only 20$\%$ of the initial value by the application of pressure. 
Thus, it is found that the rapid increase in $T_{\rm c}$ above $\sim$1.5 GPa 
accompanies a significant decrease in superconducting volume fraction in the FeSe superconductor. 
The decrease in volume fraction above 1.5 GPa was observed regardless of the use of Daphne 7373 or Daphne 7474, 
indicating that the behavior is not due to the pressure distribution 
arising from the solidification of Daphne 7373 in the sample space. 
In addition, our preliminary observation of AC susceptibility ($\chi$) using an indenter-type cell ($P$$\leq$3 GPa) 
also revealed the broadening of the transition and the rapid increase in $T_{\rm c}$ above 1.5 GPa. 
Furthermore, the decrease in volume fraction is also inferred from the 
$\rho$($T$) data above $P$=2 GPa showing the occurrence of resistivity drop 
over a wide temperature range\cite{masaki}. 
The recent X-ray experiments focused on the pressure evolution of 
the crystal structure and bonding have revealed that no structural 
transition is detected below 9 GPa, but 
a sharp decrease in Se height from the Fe plane is observed\cite{margadonna}.   
The relationship between the structural modification and 
the rapid $T_{\rm c}$ increase above 1.5 GPa has not been clarified. 
Recently, Imai $et$ $al$. have performed $^{77}$Se NMR investigations of FeSe and 
observed the disappearance of paramagnetic NMR signals below a peak of 1/$T_1$$T$ 
as a typical signature of a magnetic phase transition or spin freezing\cite{imai}. 
The decrease in superconducting volume fraction above 1.5 GPa 
is considered to be caused by the electronic inhomogeneity realized such as that in the vicinity of the magnetic phase. 
At the pressures above $P$=3 GPa, a kink is observed 
in the $\rho$$-$$T$ curve at $T_{\rm X}$$\geq$50 K\cite{masaki}. 
The anomaly at $T_{\rm X}$ may indicate a magnetic phase transition under pressure, 
although no static magnetic ordering is evidenced from the M$\ddot{\rm o}$ssbauer spectroscopy\cite{medvedev}. 
In the FeSe superconductor, there is an intriguing possibility that 
the superconducting phase is divided into two regions in the $T$$-$$P$ phase diagram, 
where an electronically inhomogeneous superconducting state is realized above 1.5 GPa. 
To elucidate it, further intensive studies are desired above $P$=1.5 GPa. 
 
In summary, it has been found that the $T_{\rm c}$$-$$P$ curve 
both for FeSe$_{1.0}$ and FeSe$_{0.8}$ exhibits a local maximum of 
$T_{\rm c}$$\sim$11 K at $P$$\sim$1 GPa, then shows 
an abrupt increase with an extremely large pressure coefficient 
for 1.5$<$$P$$<$3 GPa, and finally becomes constant for $P$$>$3 GPa. 
The maximum of $T_{\rm c}$ for $P$$>$3 GPa is $\sim$25 K (21 K) 
for FeSe$_{1.0}$ (FeSe$_{0.8}$). 
A remarkable feature is the decrease in superconducting 
volume fraction accompanied by an abrupt increase in $T_{\rm c}$ above 1.5 GPa. 
The decrease in volume fraction is considered as a signature of the electronic inhomogeneity 
in the superconduting state above 1.5 GPa. 

\acknowledgements
This work is financially supported by a Grant-in-Aid for 
Scientific Research (No. 20540355) from the Japanese Ministry of Education, 
Culture, Sports, Science and Technology.












\end{document}